\documentclass[a4paper,11pt]{article}
\usepackage{pos}

\makeatletter
\def\@preprint{\@empty}
\newcommand{\preprint}[1]{\gdef\@preprint{\hfill #1}}
\makeatother

\preprint{MIT-CTP/6169}

\title{A Breath of Fresh Air for Moli\`ere: Elastic Scatterings in OO Collisions}
\ShortTitle{A Breath of Fresh Air for Moli\`ere: Elastic Scatterings in OO Collisions}

\author*[a]{Arjun Srinivasan Kudinoor}
\author[b]{Arthur Yi-Ting Lin}
\author[c,d]{Daniel Pablos}
\author[a]{Krishna Rajagopal}

\affiliation[a]{
  Center for Theoretical Physics --- a Leinweber Institute, Massachusetts Institute of Technology, Cambridge, MA 02139}

\affiliation[b]{
 Massachusetts Institute of Technology, Cambridge, MA, 02139
}

\affiliation[c]{Departamento de F\'isica, Universidad de Oviedo, Avda. Federico Garc\'ia Lorca 18, 33007 Oviedo, Spain}

\affiliation[d]{Instituto Universitario de Ciencias y Tecnolog\'ias Espaciales de Asturias (ICTEA), Calle de la Independencia 13, 33004 Oviedo, Spain}

\emailAdd{kudinoor@mit.edu}
\emailAdd{arthur72@mit.edu}
\emailAdd{pablosdaniel@uniovi.es}
\emailAdd{krishna@mit.edu}

\abstract{
Detecting unambiguous evidence of Rutherford-like hard scatterings between energetic jet partons and quark- and gluon-like quasiparticles in quark-gluon plasma (QGP) remains a longstanding goal of heavy-ion physics. Oxygen-oxygen (OO) collisions are a promising arena in which to detect such hard scatterings. As a jet propagates through the droplet of QGP formed in the same collision, its constituents lose energy to and excite wakes in the medium, and may scatter off quark- and gluon-like quasiparticles in QGP. In this contribution to these proceedings, we show that including rare, large-angle, $2 \rightarrow 2$ hard Moli\`ere scatterings between jet partons and medium quasiparticles in the Hybrid Model is essential to reproduce recent CMS measurements of charged-particle suppression in OO collisions. We then show that Moli\`ere scatterings broaden the Soft Drop angle $R_g$, enhancing the population of jets with $R_g \gtrsim \mathcal{O}(0.1)$ in OO collisions relative to pp collisions. We also show that Moli\`ere scatterings enhance large-angle energy-energy correlations (EECs) within jets. In both cases, we propose experimental measurements where an enhancement of the relevant OO/pp ratio above unity would constitute distinctive, model-independent, evidence of hard scattering of jet partons off QGP quasiparticles. We furthermore show how the angular scale corresponding to the Moli\`ere scattering-induced deflection of jet or medium partons is imprinted in the EEC observable for jets with radius $R_{\rm jet}\sim 0.8$ in OO collisions. These results demonstrate that jet substructure measurements in OO collisions offer a promising means of probing the quasiparticles that emerge at short distances within an otherwise strongly coupled QGP.
}

\FullConference{13th International Conference on Hard and Electromagnetic Probes of High-Energy Nuclear Collisions (HardProbes2026)\\
21-26 June 2026\\
Nashville, Tennessee, USA\\}

\begin{document}

\makeatletter
\noindent\@preprint
\makeatother

\maketitle

At length scales of order its inverse temperature and longer, quark-gluon plasma (QGP) is a strongly coupled liquid. However, when it is probed at sufficiently short distances (equivalently at sufficiently high momentum-transfer), asymptotic freedom mandates the presence of quark- and gluon-like quasiparticles. Energetic partons in jets can trigger Rutherford-like hard scatterings off these quasiparticles, making jets useful probes of the microscopic structure of the strongly coupled, hot, QCD plasma.
Detecting unambiguous evidence of such hard scatterings between energetic jet partons and QGP quasiparticles remains an unrealized, long-standing, central goal of the field~\cite{Aprahamian:2015qub}.

Hard jet-medium scatterings broaden the distribution of energy and momentum within a jet. However, in large collision systems like lead-lead (PbPb), this expected broadening is obscured by the significant confounding effects of jet energy loss and jet selection bias. Recently, calculations using the hybrid strong/weak coupling model (aka Hybrid Model) were used to demonstrate that these confounding effects are considerably smaller in smaller collision systems like oxygen-oxygen (OO), making jet substructure observables in OO collisions promising probes of hard jet-medium scatterings~\cite{Kudinoor:2026wcs}. In this contribution to these proceedings, we recapitulate and extend  these results.

\vspace{0.25em}
\noindent
{\bf The Hybrid Model.} The Hybrid Model is a theoretical framework for jet quenching in which: (i) the production of jet-initiating partons and their subsequent showering according to the DGLAP evolution equations are implemented using PYTHIA 8, (ii) each parton in the parton shower loses energy and momentum as prescribed by a holographic energy loss formula~\cite{Casalderrey-Solana:2014bpa}, whose strength is controlled by a dimensionless parameter $\kappa_{\rm sc}$, which was fit to charged-hadron and jet $R_{\rm AA}$ data in PbPb collisions as described in Ref.~\cite{Kudinoor:2026wcs}, (iii) the energy and momentum lost by each parton is deposited back into the medium, exciting hydrodynamic wakes, implemented by sampling many soft hadrons from a distribution given by a suitably modified Cooper-Frye prescription~\cite{Casalderrey-Solana:2016jvj}, (iv) jet partons undergo transverse momentum broadening through multiple soft interactions with the medium, modeled as stochastic transverse momentum kicks whose magnitude is controlled by a dimensionless parameter $K$~\cite{Casalderrey-Solana:2016jvj}, and (v) jet partons can undergo rare, large-angle, high momentum-transfer, $2\to 2$ Moli\`{e}re scatterings off QGP quasiparticles in thermal equilibrium with the medium~\cite{Hulcher:2026dht}. 
Since soft jet-medium interactions 
are incorporated via strongly coupled energy loss, we ensure that the additional elastic scattering is hard enough that it needs to be treated at weak coupling by requiring
$|t|, |u| > a m_D^2$, where $t$ and $u$ are Mandelstam variables, $m_D^2 = \frac{3}{2}g_s^2 T^2$ is the Debye mass squared, and $a$ is a model parameter that sets the momentum-exchange threshold for Moli\`ere scattering.

When Moli\`ere scattering is included (excluded) we chose $\kappa_{\rm sc}=0.335$ ($0.37$), as described in Ref.~\cite{Kudinoor:2026wcs}. Following Ref.~\cite{Hulcher:2026dht}, we chose $K = 15$ and $a = 10$. Although we show comparisons to experimental data below, no experimental data in OO collisions was used when choosing any of our parameters. All calculations below were performed in minimum-bias (0-100\% central) OO collisions at $\sqrt{s_{\rm NN}}=5.36$~TeV. Band widths denote statistical uncertainties in all figures.

\begin{figure}[]
\centering
\begin{minipage}{0.35\textwidth}
\caption{Hybrid Model calculations of charged-hadron $R_{\rm OO}(p_T)$ without energy loss or Moli\`ere scattering (gray), with energy loss and wakes (blue), and with energy loss, wakes, and Moli\`ere scattering (red). Points show CMS data~\cite{CMS:2025bta} with statistical, systematic, and normalization errors added in quadrature.}
\label{fig:raa}
\end{minipage}\hfill
\begin{minipage}{0.63\textwidth}
\centering
\includegraphics[width=\linewidth,trim=10 0 50 18,clip]{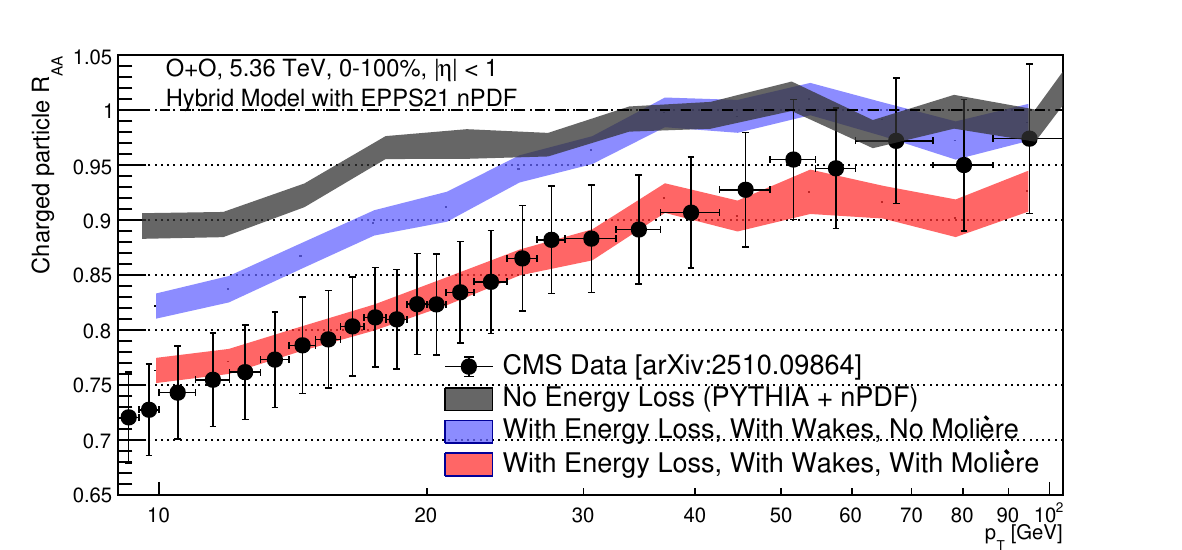}
\end{minipage}
\vspace{-0.1in}
\end{figure}

\vspace{0.25em}
\noindent
{\bf Parton Energy Loss.}
In Fig.~\ref{fig:raa}, we compare Hybrid Model calculations of charged-hadron $R_{\rm AA}$ as a function of $p_T$ with recent CMS experimental measurements~\cite{CMS:2025bta}. The gray band, which includes only initial-state effects from EPPS21 nuclear parton distribution functions (nPDFs) and excludes all effects of quenching fails to describe the CMS data. Adding strongly coupled energy loss and jet wakes (blue) yields modest suppression. Agreement with the CMS data is achieved only when the effects of Moli\`ere scattering are also included (red). This constitutes model-dependent evidence for the importance of hard scattering of jet partons off QGP quasiparticles in OO collisions.

Although Moli\`ere scatterings occur in both PbPb and OO collisions, a model-independent scaling argument indicates that their relative importance is greater in OO collisions. 
Collisional energy loss from weakly coupled elastic scatterings (e.g. Moli\`ere) scales linearly with the in-medium path length $L$ traversed by a jet parton, whereas parton energy loss scales as $L^2$ in models with weakly coupled radiative energy loss and (at least) as $L^3$ at strong coupling, as in the Hybrid Model.
Consequently, in the smaller QGP droplets produced in OO collisions, hard, elastic scatterings contribute a larger fraction of charged-hadron suppression relative to energy loss.
This makes OO collisions particularly well suited to isolating and quantifying the effects of hard scatterings between jet partons and QGP quasiparticles.

\vspace{0.25em}
\noindent
{\bf Soft Drop Angle.}
The Soft Drop grooming algorithm~\cite{Larkoski:2014wba} identifies the first hard splitting in a jet, as well as the angle $R_g$ between the two prongs involved in that splitting, while grooming away most soft hadrons originating from the wakes excited by the jet. Moli\`ere scatterings impart sizable momentum kicks to jet partons and the medium constituents they strike, increasing the angular separation of resolved splittings within jets. This should, therefore, broaden the $R_g$ distribution of jets in medium relative to that of jets in vacuum.

In PbPb collisions, however, the broadening induced by Moli\`ere scatterings is obscured by the effects of a selection bias induced by jet energy loss. Because the spectrum of jets falls steeply with increasing $p_T$, a sample of jets within a given range of reconstructed $p_T$ is biased towards those jets that lost the least amount of energy to the plasma. Such jets tend to have fewer and narrower splittings, resulting in a narrowing of the $R_g$ distribution of jets in medium~\cite{ALargeIonColliderExperiment:2021mqf,Brewer:2021hmh,CMS:2024zjn}. This selection-induced narrowing that is a consequence of energy loss can overwhelm the expected broadening induced by Moli\`ere scatterings~\cite{Brewer:2021hmh,Hulcher:2026dht}. Since the smaller QGP droplets produced in OO collisions induce less jet energy loss, the associated selection bias should be reduced, making OO collisions a promising arena in which to isolate and detect the broadening due to hard jet-medium scatterings.

\begin{figure}[]
\centering
\includegraphics[width=0.435\textwidth,trim=0 0 48 28,clip]{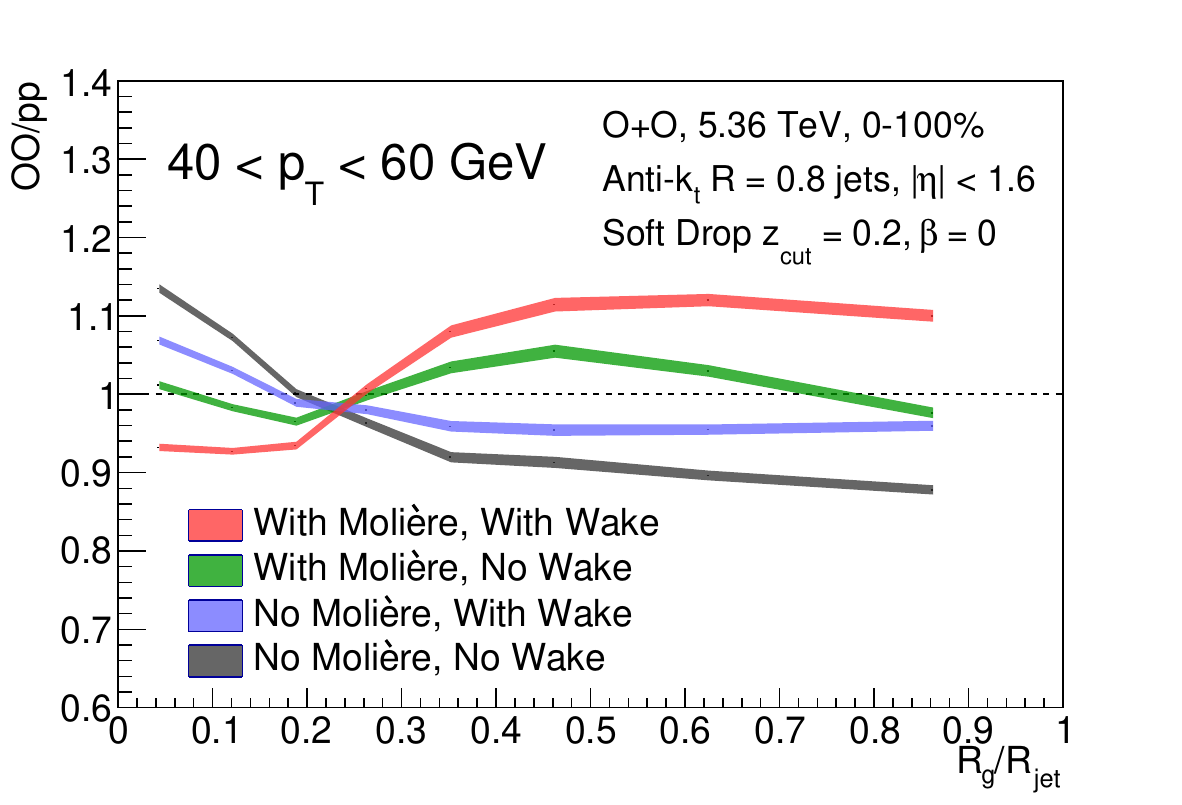}\hspace{0.025\textwidth}
\includegraphics[width=0.435\textwidth,trim=0 0 48 28,clip]{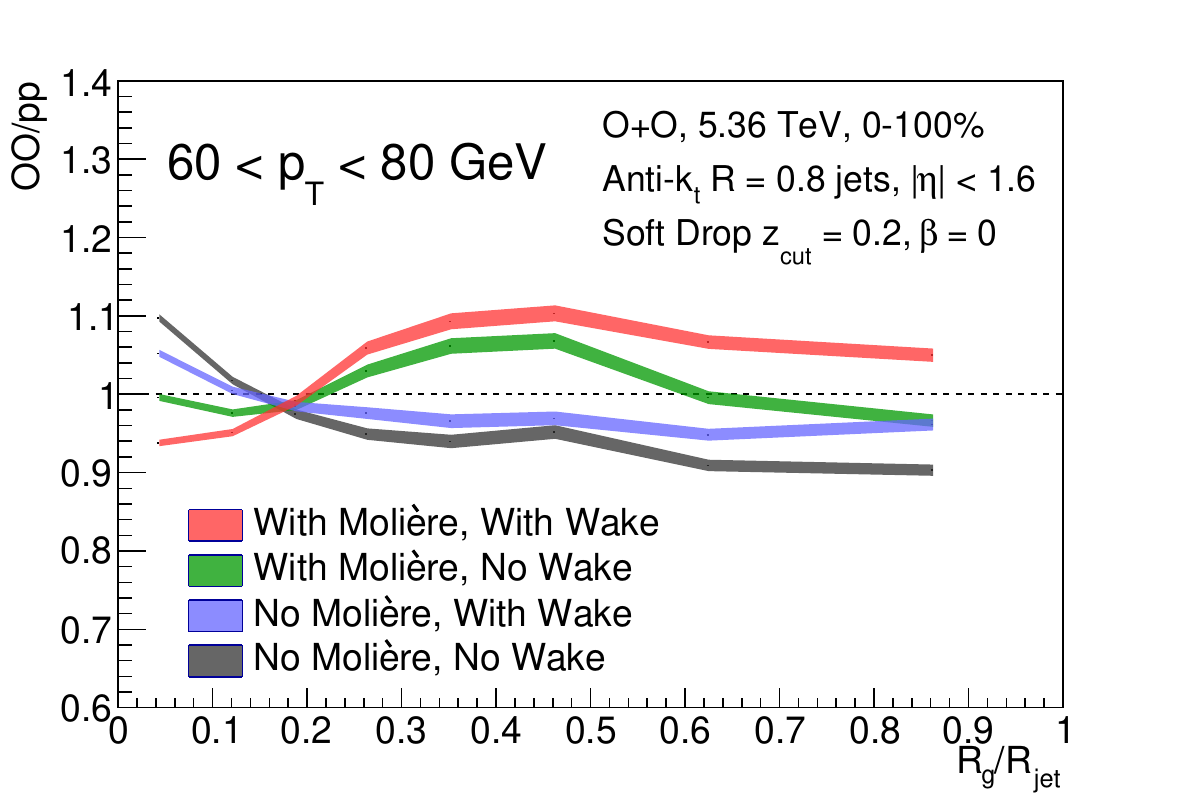}\\[2pt]
\includegraphics[width=0.435\textwidth,trim=0 0 48 28,clip]{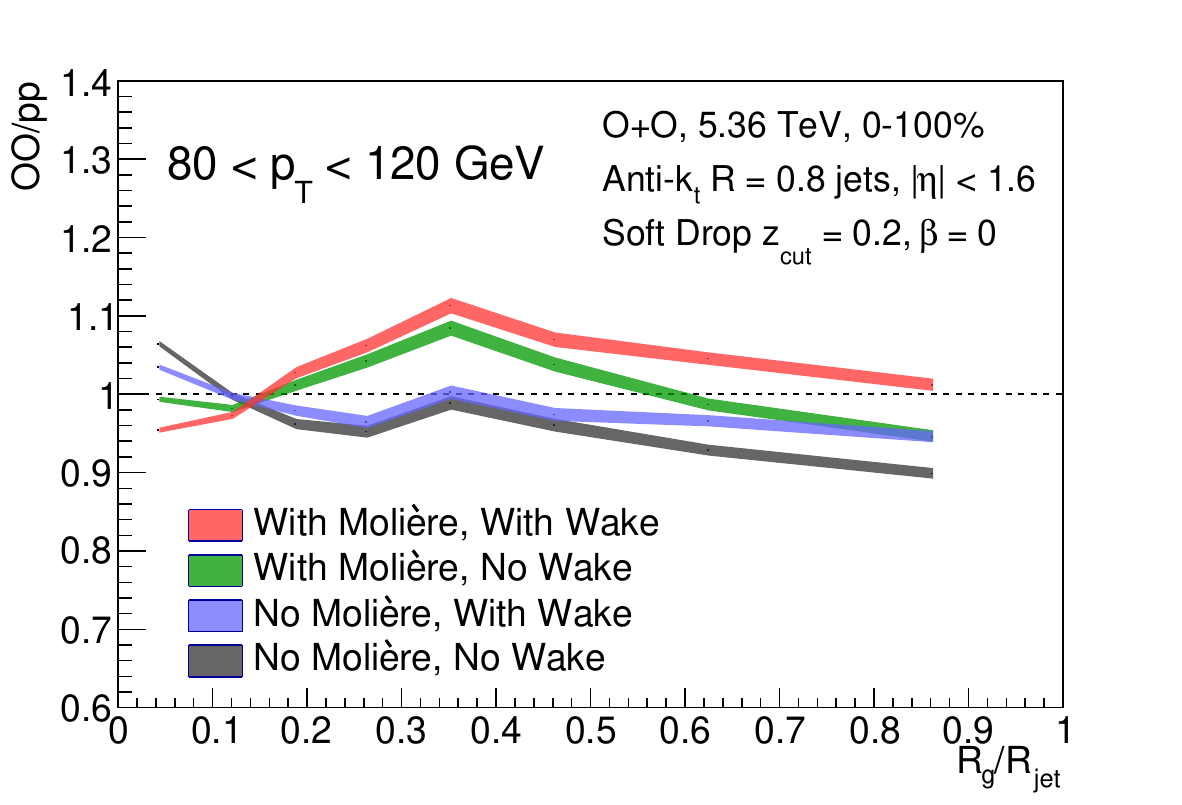}\hspace{0.025\textwidth}
\includegraphics[width=0.435\textwidth,trim=0 0 48 28,clip]{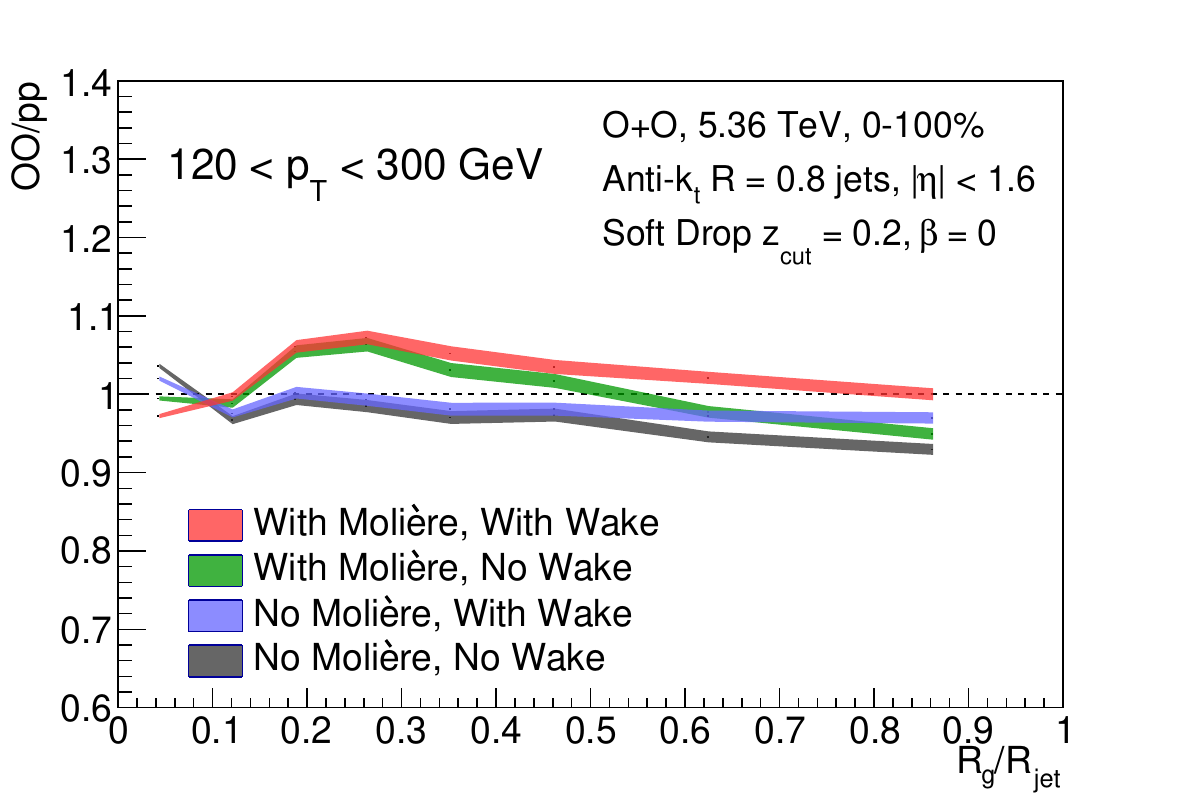}
\caption{OO/pp ratios of the $R_g$/$R_{\rm jet}$ distributions in $R_{\rm jet} = 0.8$ jets, calculated using Soft Drop grooming parameters $z_{\rm cut} = 0.2$ and $\beta = 0$. Each panel corresponds to a different range of jet $p_T$. Individual OO and pp distributions are normalized by the number of jets selected.}
\label{fig:rg}
\vspace{-0.1in}
\end{figure}

Fig.~\ref{fig:rg} shows Hybrid Model calculations of the ratio of the number of jets with a specified scaled Soft Drop angle $R_g/R_{\rm jet}$ in OO collisions to that in pp collisions for $R = 0.8$ jets of varied $p_T$ from 40 to 300 GeV. While the black and blue bands (no Moli\`ere scattering) show a modest (compared to those in PbPb collisions~\cite{Hulcher:2026dht}) narrowing of the $R_g/R_{\rm jet}$ distribution due to energy loss-induced selection bias, the green and red bands in all panels show that including Moli\`ere scattering enhances the population of jets with $R_g\gtrsim \mathcal{O}(0.1)$ in OO relative to pp collisions.
We also note that as jet $p_T$ increases, the onset of the Moli\`ere scattering-induced enhancement (where the OO/pp ratio goes above 1) shifts to smaller $R_g/R_{\rm jet}$, reflecting the fact that as the momentum of the incident jet parton in a Moli\`ere scattering increases, the typical angle between the two outgoing partons decreases~\cite{Hulcher:2026dht}.

In Ref.~\cite{Kudinoor:2026wcs}, we show analogous results for $R = 0.4$ and $R = 0.8$ jets chosen with different kinematic choices and a different Soft Drop parameter of $z_{\rm cut} = 0.1$. Our Hybrid Model calculations, both here and in Ref.~\cite{Kudinoor:2026wcs}, demonstrate that the OO/pp ratio of $R_g$ distributions can exceed unity at sufficiently large $R_g$. In inclusive jet populations in PbPb collisions, this enhancement is overwhelmed by the countervailing effects of energy loss-induced selection bias~\cite{Hulcher:2026dht}; this is not so in OO collisions! (Ref.~\cite{Hulcher:2026dht} shows that judicious selections of photon-tagged jet populations can mitigate this bias, revealing the broadening induced by Moli\`ere scatterings when jets with substantially lower $p_T$ than the tagging photon are included.) Experimental observation of such an enhancement in OO collisions would provide distinctive, model-independent evidence of hard scatterings between jet partons and QGP quasiparticles.

\vspace{0.25em}
\noindent
{\bf Energy-Energy Correlators (EECs).}
In Ref.~\cite{Kudinoor:2026wcs} and here, we also study the two-point correlator
${\rm EEC}(R_L)
\equiv \frac{1}{\mathcal{N}}\,\frac{1}{\delta r}
\sum_{\rm jets}
\, \, \sum_{{\rm pairs} 
} 
p_{T,i}\, p_{T,j}
$, following the definition, jet reconstruction procedure, and track identification procedure of Ref.~\cite{CMS:2025ydi}.
Here, $(i,j)$ refers to pairs of charged particles that are both within an angular distance $R_{\rm jet}$ of the winner-take-all axis of E-scheme jets. 
$R_L \equiv \sqrt{(\eta_i - \eta_j)^2 + (\phi_i - \phi_j)^2}$ is the angular separation between particles $i$, $j$, the sum is taken only over pairs whose $R_L$ falls within a bin of width $\delta r$, and $\mathcal{N}$ is chosen such that $\int_{0.008}^{R_{\rm jet}}{\rm EEC}(R_L)\, d R_L = 1$.

\begin{figure}[]
\centering
\includegraphics[width=0.465\textwidth,trim=0 0 24 10,clip]{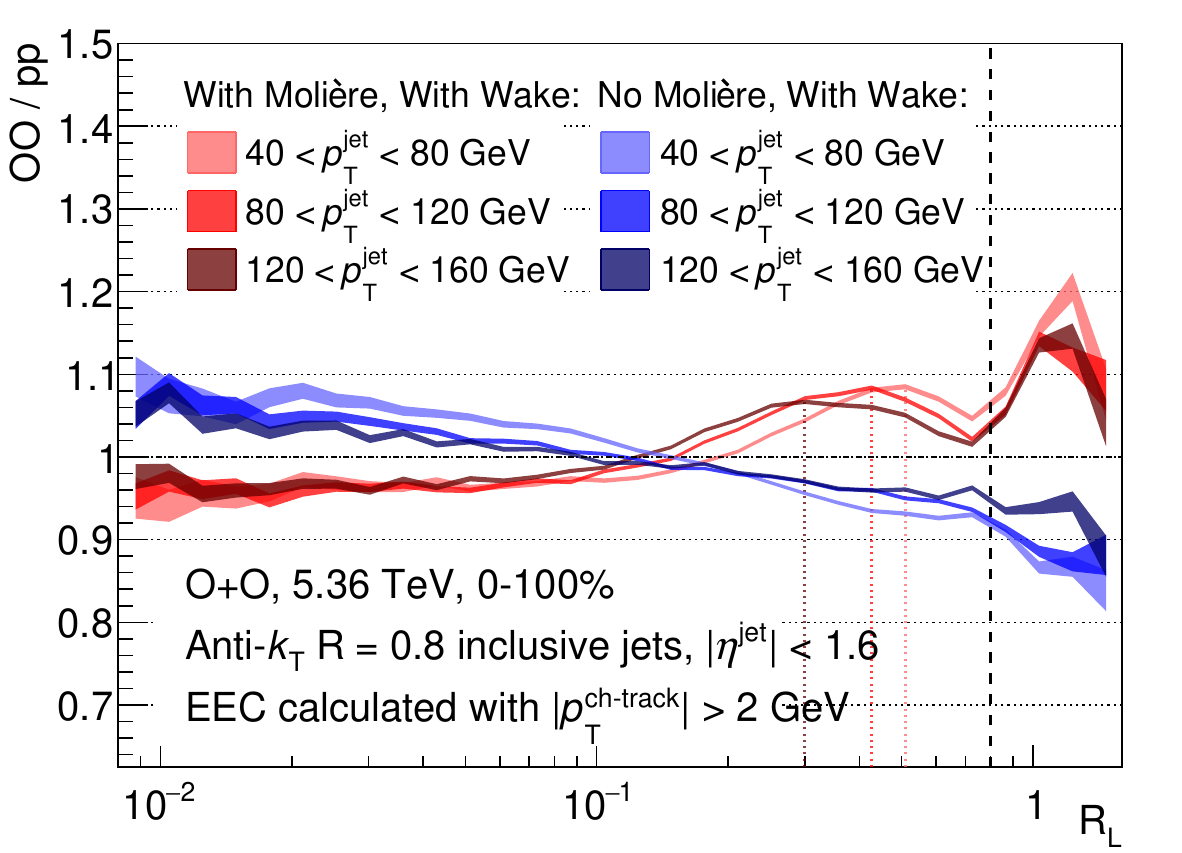}\hspace{0.025\textwidth}
\includegraphics[width=0.465\textwidth,trim=0 0 24 10,clip]{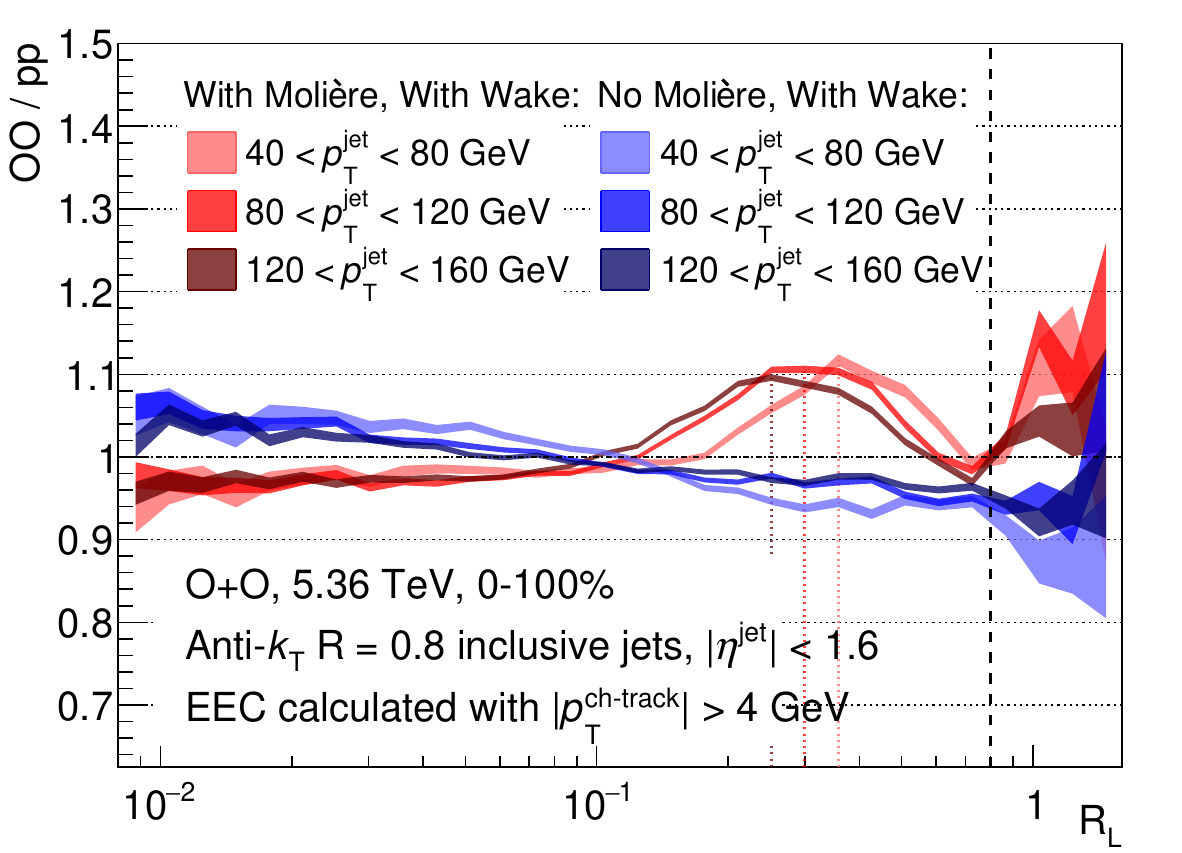}
\caption{OO/pp EEC ratios for $R_{\rm jet}=0.8$ jets with $|\eta^{\rm jet}|<1.6$, using charged tracks with $p_T>2$~GeV (left) and $p_T>4$~GeV (right). Red (blue) includes (excludes) Moli\`ere scattering. Darker shades denote higher jet $p_T$. The dashed vertical lines mark $R_L=R_{\rm jet}$, while the dotted red lines mark local maxima below $R_{\rm jet}$.}
\label{fig:eec}
\vspace{-0.1in}
\end{figure}

Fig.~\ref{fig:eec} shows Hybrid Model calculations of the OO/pp ratios of EECs for $R_{\rm jet}=0.8$ jets in three jet $p_T$ ranges, calculated using charged-particles with $p_T>2$ GeV (left) and $p_T>4$ GeV (right) (chosen to remove the soft hadrons originating from jet wakes). The blue bands (no Moli\`ere scattering) show a modest enhancement (suppression) of small-$R_L$ (large-$R_L$) correlations in OO relative to pp collisions, arising from the energy loss-induced bias towards selecting jets with narrower hard splittings. With Moli\`ere scattering (red bands), the OO/pp EEC ratios exceed unity for $0.2\lesssim R_L<R_{\rm jet}$, reflecting jet substructure broadening due to hard jet-medium scatterings.

Furthermore, this large-$R_L$ enhancement has a striking shape: the red bands exhibit a distinct local maximum, or bump, at $R_L \sim 0.3-0.5$, which appears for jets with radius large enough to capture the scattered particles~\cite{Kudinoor:2026wcs} and whose position reflects a typical scattering angle associated with Moli\`ere scattering, as discussed further in the companion contribution~\cite{Kudinoor:2026xxx} to these proceedings.
As jet $p_T$ increases, the bump shifts to smaller $R_L$, consistent with the fact that as the momentum of the incident jet parton in a Moli\`ere scattering increases, the typical angle between the two outgoing partons decreases~\cite{Hulcher:2026dht,Kudinoor:2026xxx}. Since the momentum spectrum of the soft hadrons originating from jet wakes has not yet been determined quantitatively from experimental data, 
and since hydrodynamic calculations indicate that they may contribute up to 
$p_T$ $\sim$ $2.5$~GeV~\cite{Casalderrey-Solana:2026ftw} (cf.~in the simplified treatment implemented in the Hybrid Model~\cite{Casalderrey-Solana:2016jvj} they are insignificant for $p_T\gtrsim 1.5$~GeV),
it is important to establish that the distinctive contribution from Moli\`ere scattering 
seen in Fig.~\ref{fig:eec}
is robust against variations in the minimum charged-particle $p_T$ used to calculate the EEC. In the right panel of Fig.~\ref{fig:eec}, we see that both the enhancement above unity and the $p_T$-dependent bump persist when the minimum charged-particle $p_T$ is raised to $4$ GeV, indicating that the signatures of Moli\`ere scattering are robust against a restrictive track selection that should fully eliminate contributions from wakes.

Experimental observation of an OO/pp ratio of EECs exceeding unity at large $R_L$ would provide distinctive, model-independent, evidence of hard scatterings between jet partons and QGP quasiparticles. Furthermore, measuring a jet $p_T$-dependent bump in the OO/pp ratio of EECs could help identify the characteristic angles at which partons are scattered. Ref.~\cite{Kudinoor:2026wcs} presents additional predictions for different jet radii and charged-particle $p_T$ cuts, while the companion contribution~\cite{Kudinoor:2026xxx} to these proceedings present model-independent pQCD calculations that mandate the presence of this bump and describe its dependence on jet $p_T$, local medium temperature, and the value of the threshold parameter $a$ in the Hybrid Model treatment of Moli\`ere scatterings.

\begingroup
\noindent

\vspace{0.25em}
\noindent
{\bf Acknowledgments.}
Research supported in part by the U.S.~DOE, Office of Science, Office of Nuclear Physics grant 
DE-SC0011090, by Ram\'on y Cajal fellowship RYC2023-044989-I, by the NSF Graduate Research Fellowship Program grant~2141064, and by subMIT at MIT Physics~\cite{Acosta:2025kgk}.


\par\endgroup
\bibliographystyle{JHEP}
\bibliography{bibliography}

\end{document}